\documentclass[prl,twocolumn,noshowkeys,noshowpacs]{revtex4}%
\usepackage{amsfonts}
\usepackage{amsmath}
\usepackage{amssymb}
\usepackage{graphicx}%
\begin{document}
\title{Antiferromagnetic fluctuations and d-wave superconductivity in electron-doped
high-temperature superconductors}
\author{Bumsoo Kyung$^{1}$, Jean-S\'{e}bastien Landry$^{1}$, and A.-M.S.
Tremblay$^{1,2}$}
\affiliation{$^{1}$D\'{e}partement de physique and Regroupement qu\'{e}b\'{e}cois sur les
mat\'{e}riaux de pointe, $^{2}$Institut canadien de recherches avanc\'{e}es,
Universit\'{e} de Sherbrooke, Sherbrooke, Qu\'{e}bec, Canada, J1K 2R1}
\date{\today}

\begin{abstract}
We show that, at weak to intermediate coupling, antiferromagnetic fluctuations
enhance d-wave pairing correlations until, as one moves closer to
half-filling, the antiferromagnetically-induced pseudogap begins to suppress
the tendency to superconductivity. The accuracy of our approach is gauged by
detailed comparisons with Quantum Monte Carlo simulations. The negative
pressure dependence of $T_{c}$ and the existence of photoemission hot spots in
electron-doped cuprate superconductors find their natural explanation within
this approach.

\end{abstract}
\maketitle

For almost two decades, the mechanism for high temperature superconductivity
has been one of the main issues in condensed matter physics. Despite an
extensive body of theoretical work, there is at present no consensus about the
mechanism. This is mainly due to lack of a reliable theoretical tool for a
strong coupling problem where the value of the on-site Coulomb interaction $U$
is of the order of, or larger than, the bandwidth\ \cite{Aeppli:2001}. The
situation, however, appears more promising for electron-doped high-temperature
superconductors (e-HTSC) in which the charge gap at half-filling is $25\%$
smaller than that of hole-doped cuprates (h-HTSC), suggesting a smaller value
of $U$\ \cite{Uchida:1989}. This offers an opportunity for theories of
$d$-wave superconductivity based on weak- to intermediate-coupling
approaches\ \cite{Bickers-Scalapino:1989,Chubukov:2002,Carbotte:1999}.

In this communication, we show that improved theoretical calculations can
indeed describe several aspects of e-HTSC that were unexplained by previous
calculations. For example, we show that the negative pressure derivative of
the superconducting transition temperature $T_{c}$ of e-HTSC, which contrasts
with the positive pressure derivative of h-HTSC, can be explained. In
addition, the hot spots observed in Angle Resolved Photoemission Experiments
(ARPES) also come out of the calculation. We also discuss how a decrease in
$T_{c}$ in the underdoped region can occur when a large pseudogap is produced
by antiferromagnetic (AFM) fluctuations. Previous
calculations\ \cite{Pao:1995} predicted that $T_{c}$ would increase
monotonically as one approaches half-filling.
%TCIMACRO{\TeXButton{Begin figure}{\begin{figure}} }%
%BeginExpansion
\begin{figure}
%EndExpansion
%TCIMACRO{\TeXButton{QMC_susceptibility.EPS}{\centerline{\includegraphics
%[width=8cm]{QMC_susceptibility.EPS}}} }%
%BeginExpansion
\centerline{\includegraphics[width=8cm]{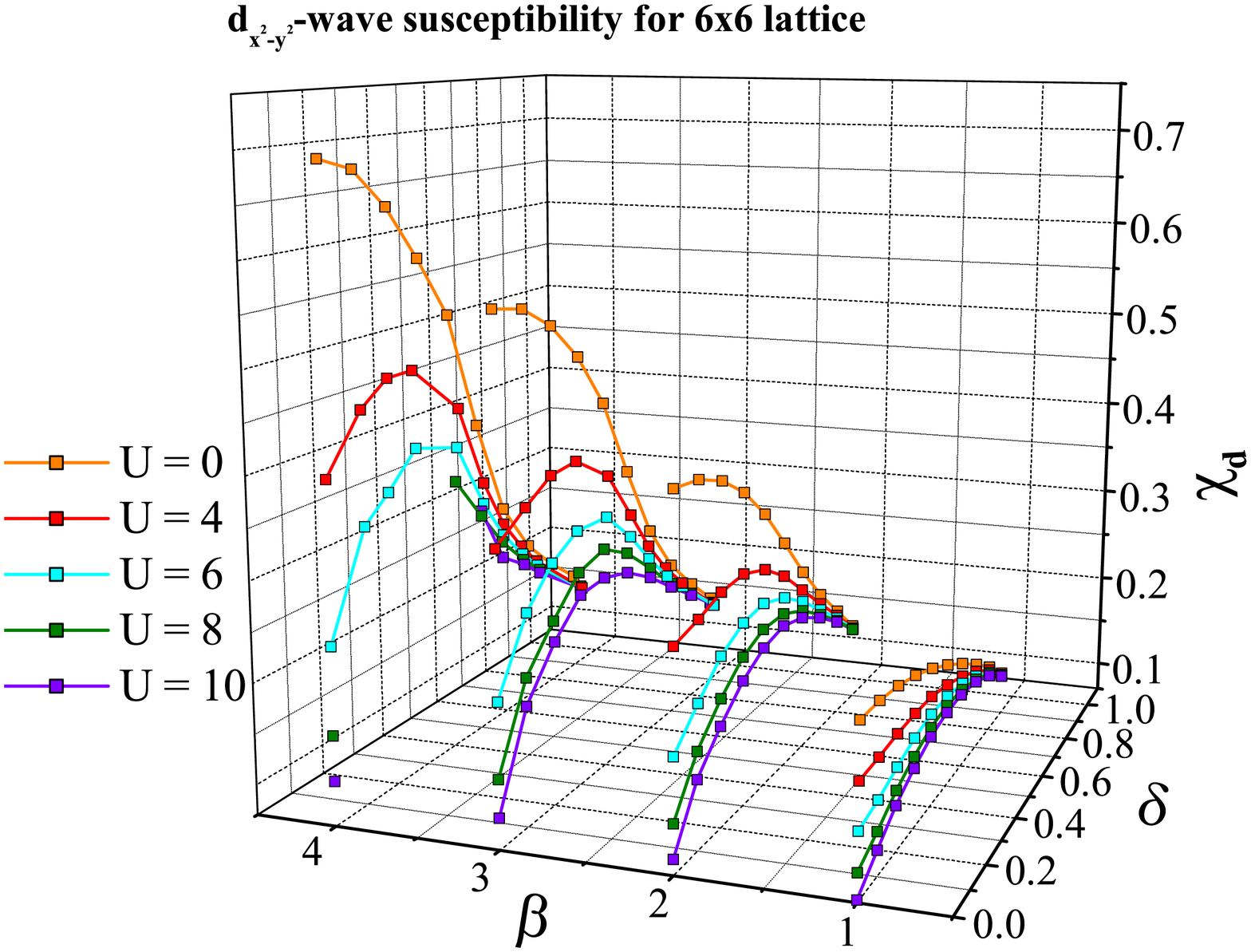}}
%EndExpansion
%TCIMACRO{\TeXButton{Caption}{\caption{(Color) The $d_{x^2-y^2}$
%susceptibility obtained from QMC simulations as a function of doping
%and of temperature for
%a $6\times6$ lattice. Various values of $U$ correspond to different colors.
%The size dependence of the results is small
%at these temperatures. The Trotter step size is $\Delta\tau=1/10$ while
%the number of measurements at each point in parameter space is around $10^5$.
%Measurements are grouped in blocs of $125$ and stabilized every five steps along
%the imaginary-time axis.}} }%
%BeginExpansion
\caption{(Color) The $d_{x^2-y^2}$
susceptibility obtained from QMC simulations as a function of doping
and of temperature for
a $6\times6$ lattice. Various values of $U$ correspond to different colors.
The size dependence of the results is small
at these temperatures. The Trotter step size is $\Delta\tau=1/10$ while
the number of measurements at each point in parameter space is around $10^5$.
Measurements are grouped in blocs of $125$ and stabilized every five steps along
the imaginary-time axis.}
%EndExpansion
%TCIMACRO{\TeXButton{fig1}{\label{fig1}} }%
%BeginExpansion
\label{fig1}
%EndExpansion
%TCIMACRO{\TeXButton{End Figure}{\end{figure}}}%
%BeginExpansion
\end{figure}%
%EndExpansion

Let us first consider the results of numerical calculations concerning the
possibility of $d$-wave superconductivity in the Hubbard model. In Fig.1 we
present a rather detailed survey of the $d_{x^{2}-y^{2}}$-wave susceptibility
$\chi_{d}$ obtained from Quantum Monte Carlo (QMC)
calculations\ \cite{Hirsch:1988,White:1989,Moreo:1991,Scalettar:1991} for the
Hubbard model. The Hubbard model is characterized, as usual, by
nearest-neighbor hopping $t$ and on-site repulsion $U.$ By contrast with
variational methods, the QMC calculations are unbiased. They also can be
performed on much larger lattices than exact diagonalizations. QMC is
essentially exact, within statistical error bars that, in Fig.\ 1, are
generally smaller than the symbol size. As usual, the $d_{x^{2}-y^{2}}$-wave
susceptibility is defined by $\chi_{d}=\int_{0}^{\beta}d\tau\left\langle
T_{\tau}\Delta\left(  \tau\right)  \Delta^{\dagger}\right\rangle $ with the
$d$-wave order parameter $\Delta^{\dagger}=\sum_{i}\sum_{\gamma}g\left(
\gamma\right)  c_{i\uparrow}^{\dagger}c_{i+\gamma\downarrow}^{\dagger}$ the
sum over $\gamma$ being over nearest-neighbors, with $g\left(  \gamma\right)
=\pm1/2$ depending on whether $\gamma$ is a neighbor along the $\widehat
{\mathbf{x}}$ or the $\widehat{\mathbf{y}}$ axis. From now on, we work in
units where $k_{B}=1,$ $\hbar=1,$ lattice spacing and hopping $t$ are unity$.$
The results are shown for various temperatures $T=\beta^{-1}$, dopings
$\delta$ and interaction strengths $U$ (shown by the various colors). The data
clearly shows that the dome shape dependence of $\chi_{d}$ is present not only
for strong coupling $(U\gtrsim8),$but also at weak to intermediate coupling
$(U=4).$ For weak coupling the dome shape occurs at temperatures that are
sufficiently low $\left(  \beta=4\right)  $ for AFM (or spin-density-wave)
correlations to build up. It has been known for a long time that these
results, obtained by a numerical method of choice, by themselves do not
suffice to decide whether there is a $d$-wave superconducting phase in the
Hubbard model. Indeed, the susceptibility should diverge if there is a phase
transition. Also, at $\beta=4$ the non-interacting model, $U=0$, has a larger
susceptibility than the $U\neq0$ model, a fact that does not encourage optimism.

To conclusively verify whether $d$-wave superconductivity exists in this model
at weak to intermediate coupling, one needs to reach temperatures that are an
order of magnitude smaller than those shown in Fig.1. As is well known, the
so-called sign-problem renders impossible simulations at these low
temperatures. To reach such temperatures, we use the Two-Particle
Self-Consistent approach\ \cite{Vilk:1997,Allen:2002} (TPSC) and extend it to
compute superconducting correlations. The accuracy of the method for spin
fluctuations and self-energy has already been
proven\ \cite{Vilk:1997,Moukouri:2000} by comparisons with QMC data. In
particular, there is a pseudogap of AFM origin at a crossover temperature
$T_{X}.$

Briefly speaking, to extend TPSC to compute pairing susceptibility, we begin
from the Schwinger-Martin-Kadanoff-Baym formalism with both
diagonal\ \cite{Vilk:1997,Allen:2002} and off-diagonal\ \cite{Allen:2001}
source fields. The self-energy is expressed in terms of spin and charge
fluctuations and the irreducible vertex entering the Bethe-Salpeter equation
for the pairing susceptibility is obtained from functional differentiation.
The final expression for the $d$-wave susceptibility is,
%TCIMACRO{\TeXButton{%
%\begin{widetext}
%}{\begin{widetext}} }%
%BeginExpansion
\begin{widetext}
%EndExpansion
\begin{align}
\chi_{d}\left(  \mathbf{q}=0,iq_{n}=0\right)   &  =\frac{T}{N}\sum_{k}\left(
g_{d}^{2}\left(  \mathbf{k}\right)  G_{\uparrow}^{\left(  2\right)  }\left(
-k\right)  G_{\downarrow}^{\left(  2\right)  }\left(  k\right)  \right)
-\frac{U}{4}\left(  \frac{T}{N}\right)  ^{2}\sum_{k,k^{\prime}}g_{d}\left(
\mathbf{k}\right)  G_{\uparrow}^{\left(  2\right)  }\left(  -k\right)
G_{\downarrow}^{\left(  2\right)  }\left(  k\right) \nonumber\\
&  \times\left(  \frac{3}{1-\frac{U_{sp}}{2}\chi_{0}\left(  k^{\prime
}-k\right)  }+\frac{1}{1+\frac{U_{ch}}{2}\chi_{0}\left(  k^{\prime}-k\right)
}\right)  G_{\uparrow}^{\left(  1\right)  }\left(  -k^{\prime}\right)
G_{\downarrow}^{\left(  1\right)  }\left(  k^{\prime}\right)  g_{d}\left(
\mathbf{k}^{\prime}\right)  . \label{Suscep_d}%
\end{align}%
%TCIMACRO{\TeXButton{%
%\end{widetext}
%}{\end{widetext}}}%
%BeginExpansion
\end{widetext}%
%EndExpansion
In the above expression, Eq.(\ref{Suscep_d}), $g_{d}\left(  \mathbf{k}\right)
$ is the form factor for the gap symmetry, while $k$ and $k^{\prime}$ stand
for both wave-vector and Matsubara frequencies $k\equiv\left(  \mathbf{k}%
,\left(  2n+1\right)  \pi T\right)  $ on a square-lattice with $N$ sites at
temperature $T.$ The spin and charge susceptibilities take the form $\chi
_{sp}^{-1}\left(  q\right)  =\chi_{0}(q)^{-1}-\frac{U_{sp}}{2}$ and $\chi
_{ch}^{-1}\left(  q\right)  =$ $\chi_{0}(q)^{-1}+\frac{U_{ch}}{2}$ with
$\chi_{0}$ computed with the Green function $G_{\sigma}^{(1)}$ that contains
the self-energy whose functional differentiation gave the spin and charge
vertices. The values of $U_{sp},$ $U_{ch}$ and $\left\langle n_{\uparrow
}n_{\downarrow}\right\rangle $ are obtained\ \cite{VilkChen:1994} from
$U_{sp}=U\left\langle n_{\uparrow}n_{\downarrow}\right\rangle /\left(
\left\langle n_{\uparrow}\right\rangle \left\langle n_{\downarrow
}\right\rangle \right)  $ and from the local-moment sum-rule. In the pseudogap
regime, one cannot use $U_{sp}=U\left\langle n_{\uparrow}n_{\downarrow
}\right\rangle /\left(  \left\langle n_{\uparrow}\right\rangle \left\langle
n_{\downarrow}\right\rangle \right)  $. Instead \cite{Vilk:1997}, one uses the
local-moment sum rule with the zero temperature value of $\left\langle
n_{\uparrow}n_{\downarrow}\right\rangle \ $obtained by the method of
Ref.\ \cite{Hanke:1990} that agrees very well with QMC calculations at all
values of $U.$ Also, $G_{\sigma}^{(2)}$ contains self-energy effects coming
from spin and charge fluctuations, as described in previous
work\ \cite{Moukouri:2000,Allen:2002}. \ 

The effective interaction in the particle-particle channel mediated by AFM
fluctuations is represented by the second term of Eq.(\ref{Suscep_d}). It
becomes sizeable only after spin fluctuations have become large.
Eq.(\ref{Suscep_d}) thus contains two leading effects, namely spin and charge
fluctuations influence the magnitude of the effective interactions in the
particle-particle channel and they also decrease the lifetime of particles
that pair (through $G_{\sigma}^{(2)}$). The latter effect is generally
detrimental to superconductivity while the former may favor pairing.

The explicit expression for $\chi_{d},$ Eq.(\ref{Suscep_d}), allows us to find
\textit{analytically} which gap symmetry is enhanced or suppressed by AFM
fluctuations. Indeed, since near half-filling AFM fluctuations are strongly
peaked at $\mathbf{k}^{\prime}-\mathbf{k=Q}$ (commensurate or incommensurate),
the sign of $f\equiv-g_{d}\left(  \mathbf{k+Q}\right)  /g_{d}\left(
\mathbf{k}\right)  $ and the magnitude of $g_{d}\left(  \mathbf{k}\right)  $
near the Fermi wave vector $\mathbf{k}_{F}$ determine the most favorable gap
symmetry. Within a spin-singlet subspace, $s$-wave and $d_{xy}$-wave
symmetries are suppressed since $f<0$. Extended $s$-wave symmetry has $f>0,$
just like $d_{x^{2}-y^{2}}$-wave$,$ but its form factor is much smaller near
$\mathbf{k}_{F}$, so we take $g_{d}\left(  k\right)  =\left(  \cos k_{x}-\cos
k_{y}\right)  $.

Let us first verify the accuracy of this approach by comparing, in Fig.2, the
QMC results for $\chi_{d}$, shown by symbols, with those of the generalized
TPSC approach, Eq.(\ref{Suscep_d}), indicated by the solid line. The case
$U=0,$ $\beta=4$ is for reference. Fig. 2 demonstrates that the approach,
Eq.(\ref{Suscep_d}), agrees very well with QMC results for $\chi_{d}$ at
$U=4$. The agreement improves for lower values of $U$. When the interaction
strength reaches the intermediate coupling regime, $U=6$, deviations of the
order of $20$ to $30\%$ may occur but the qualitative dependence on
temperature and doping remains accurate. The inset shows that previous
spin-fluctuation calculations (FLEX) in two
dimensions\ \cite{Bickers-Scalapino:1989,Pao:1995} deviate both qualitatively
and quantitatively from the QMC results. More specifically, in the FLEX
approach $\chi_{d}$ does not show a pronounced maximum at finite doping.
Moreover, it is known from previous work that FLEX does not show a pseudogap
in the single-particle spectral weight at the Fermi
surface\ \cite{Moukouri:2000}. In TPSC the pseudogap is the key ingredient
that leads to a decrease in $T_{c}$ in the underdoped regime.
%TCIMACRO{\TeXButton{Begin figure}{\begin{figure}}}%
%BeginExpansion
\begin{figure}%
%EndExpansion%
%TCIMACRO{\TeXButton{Comp_MCQ.EPS}{\centerline{\includegraphics
%[width=8cm]{Comp_MCQ.EPS}}} }%
%BeginExpansion
\centerline{\includegraphics[width=8cm]{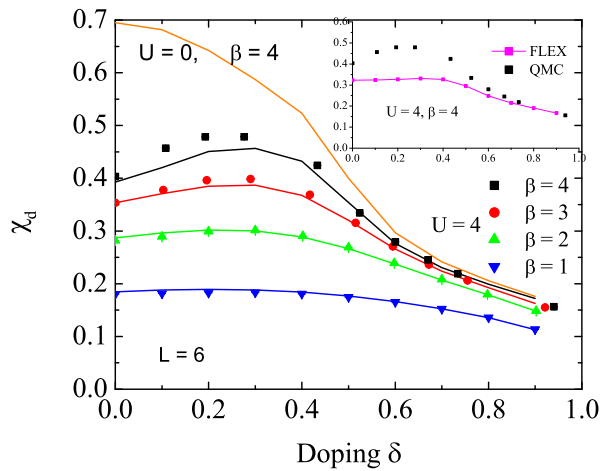}}
%EndExpansion
%TCIMACRO{\TeXButton{Caption}{\caption
%{(Color online) Comparisons between the $d_{x^2-y^2}$
%susceptibility obtained from QMC simulations (see previous figure) and from the approach
%described in the present work. QMC error bars are smaller than the symbols.
%Analytical results are joined by solid lines. Both calculations are for $U=4$,
%a $6\times6$ lattice and four different temperatures. The case $U=0, \beta=4$
%is shown for reference. The size dependence of the results is small
%at these temperatures. The inset compares QMC and FLEX at $U=4, \beta=4$.}} }%
%BeginExpansion
\caption{(Color online) Comparisons between the $d_{x^2-y^2}$
susceptibility obtained from QMC simulations (see previous figure) and from the approach
described in the present work. QMC error bars are smaller than the symbols.
Analytical results are joined by solid lines. Both calculations are for $U=4$,
a $6\times6$ lattice and four different temperatures. The case $U=0, \beta=4$
is shown for reference. The size dependence of the results is small
at these temperatures. The inset compares QMC and FLEX at $U=4, \beta=4$.}
%EndExpansion
%TCIMACRO{\TeXButton{fig2}{\label{fig2}} }%
%BeginExpansion
\label{fig2}
%EndExpansion
%TCIMACRO{\TeXButton{End Figure}{\end{figure}}}%
%BeginExpansion
\end{figure}%
%EndExpansion

In TPSC we can understand why, as mentioned above, $\chi_{d}$ is smaller than
the non-interacting value in this temperature range. Indeed, the main
contribution is from the first term in Eq.(\ref{Suscep_d}) which represents a
pair of propagating particles that do not interact with each other. The
contribution of the second term, which represents interaction through the
exchange of spin and charge fluctuations, is, for $\beta=4$, about $1\%$ at
$\delta=0.5$, growing to only $22\%$ at $\delta=0$. Hence, in this temperature
range, $\chi_{d}$ is smaller than the non-interacting value because of the
decrease in spectral weight at $\omega=0$ brought about by AFM self-energy effects.

While it is impossible to do QMC calculations at lower $T$, the analytical
formula for $\chi_{d}$, Eq.(\ref{Suscep_d}), can be extended to low $T$ and to
$256\times256$ lattice size using renormalization group
acceleration\ \cite{PaoBickers:1994} and Fast Fourier Transforms. This allows
us to verify whether there is $d$-wave superconductivity ($d-SC)$ in the
Hubbard model at weak to intermediate coupling. The complete Bethe-Salpeter
equation would contain the possibility of repeatedly exchanging spin
fluctuations. Eq.(\ref{Suscep_d}) contains only the first two terms, namely
the zero and the one spin- and charge fluctuation exchange. As for the
expansion $\left(  1-x\right)  ^{-1}\sim1+x,$ the divergence should occur when
the first two terms have the same magnitude. We can thus estimate $T_{c}$ for
$d-SC.$ As usual, the $T_{c}$ obtained from the divergence of the infinite
series (Thouless criterion) should give an upper bound to the
Kosterlitz-Thouless transition temperature $T_{KT}$ expected in $d=2$. In
Fig.3 (a) the first (DOS) and second (Vertex) contributions in
Eq.(\ref{Suscep_d}) are plotted for $U=4$ at $\beta=64$ as a function of
doping. The vertex part becomes larger than the first part over a range of
$\delta.$ This signals, according to our criterion, that $0.07<\delta<0.13$ is
below $T_{c}$ at $\beta=64$. Note that it is because the vertex part decreases
much faster than the DOS part near half-filling that the $d-SC$ stops close to
half-filling, leading to a dome shape in $T_{c}$. The fast decrease of the
vertex part near half-filling is because it has its strongest contribution
near the Fermi surface where the pseudogap effect is most pronounced.%
%TCIMACRO{\TeXButton{Begin figure}{\begin{figure}} }%
%BeginExpansion
\begin{figure}
%EndExpansion
%TCIMACRO{\TeXButton{Tc.EPS}{\centerline{\includegraphics[width=7cm]{Tc.EPS}}}
%}%
%BeginExpansion
\centerline{\includegraphics[width=7cm]{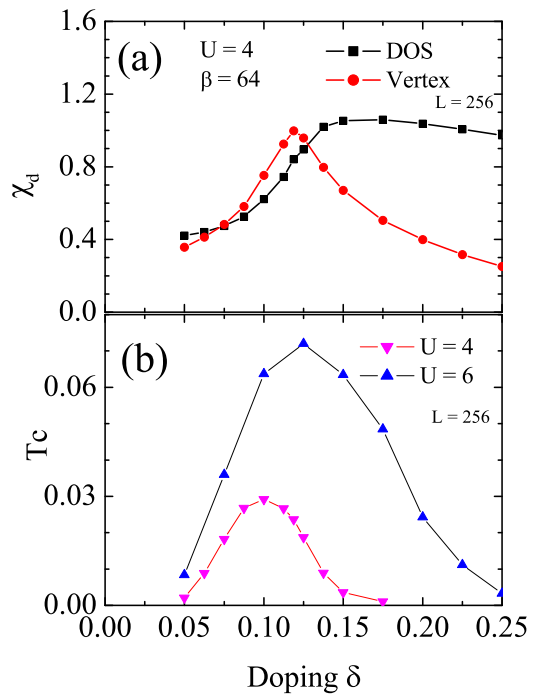}}
%EndExpansion
%TCIMACRO{\TeXButton{Caption}{\caption
%{(Color online) Part (a) shows the contributions from the first term (DOS) and second term (vertex)
%of Eq.(\protect\ref{Suscep_d}). In  (b), our estimate of $T_c$ using the
%Thouless criterion for $U=4$ and $U=6$.}} }%
%BeginExpansion
\caption
{(Color online) Part (a) shows the contributions from the first term (DOS) and second term (vertex)
of Eq.(\protect\ref{Suscep_d}). In  (b), our estimate of $T_c$ using the
Thouless criterion for $U=4$ and $U=6$.}
%EndExpansion
%TCIMACRO{\TeXButton{fig3}{\label{fig3}} }%
%BeginExpansion
\label{fig3}
%EndExpansion
%TCIMACRO{\TeXButton{End Figure}{\end{figure}}}%
%BeginExpansion
\end{figure}%
%EndExpansion

The solid lines with symbols in Fig.\ 3(b) give the value of $T_{c}$ estimated
for two values of $U$ in the intermediate coupling regime. The $U=6$ results
should be viewed as giving the qualitative trend with increasing $U.$ As is
clear by now, the decrease of $T_{c}$ near half-filling is caused by the same
AFM fluctuations that enhance $d-SC$ at large doping. $d-SC$ fluctuations in
our approach are important only between $T_{c}$ and $T_{KT}$, by contrast with
phase fluctuation theories at strong coupling\ \cite{EmeryKivelson:1995}. Our
results also contrast with theories where the decrease of $T_{c}$ is driven by
hidden competing broken symmetry\ \cite{Chakravarty:2000}.

To make more detailed connection with experimental results on e-HTSC, one
should add second-neighbor $t^{\prime}$ and third-neighbor $t^{\prime\prime}$
hopping to the Hamiltonian, as suggested by band-structure calculations and by
ARPES. We perform the usual particle-hole transformation that maps
electron-doping of the negative $t^{\prime}$ model to hole doping with
positive $t^{\prime}.$ As $t^{\prime}$ and $t^{\prime\prime}$ increase, AFM
fluctuations are frustrated, so pseudogap effects become less important and
the fall of $T_{c}$ on the underdoped side becomes less and less pronounced.
Including AFM coupling in the third dimension would lead to a real AFM
transition that would eventually overcome the pairing instability. The more
significant result we want to draw attention to is that, assuming that
applying pressure only increases $t$ (and thus decreases $U/t$), the data of
Fig.\ 3 shows that $d\ln T_{c}/dP<0$ in the weak to intermediate coupling
regime described by our approach. This remains true with finite $t^{\prime}$
and $t^{\prime\prime}$ and agrees with the experimental negative pressure
dependence of $T_{c}$ in these compounds\ \cite{Maple:1990}. By contrast, in
h-HTSC $d\ln T_{c}/dP$ has the opposite sign. If antiferromagnetism plays a
role in the superconductivity of both e-HTSC and h-HTSC, then the positive
sign of $d\ln T_{c}/dP$ in the latter may be understood from the fact that
they are in the strong-coupling regime where $J=4t^{2}/U$ increases with pressure.%

%TCIMACRO{\TeXButton{Begin figure[tbp]}{\begin{figure}[tbp]} }%
%BeginExpansion
\begin{figure}[tbp]
%EndExpansion
%TCIMACRO{\TeXButton{Hot_spots.ps}{\includegraphics[width=8.5cm]{Hot_spots.ps}}
%}%
%BeginExpansion
\includegraphics[width=8.5cm]{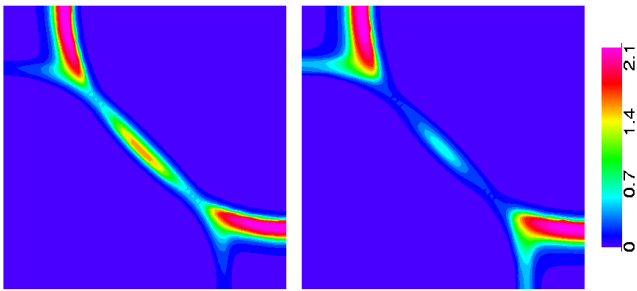}
%EndExpansion
%TCIMACRO{\TeXButton{Caption}{\caption
%{(Color) Fermi surface plots obtained from energy dispersion curves integrated from
%$-0.2$ to $0.1$ with $|t'|=0.175$, $|t''|=0.05$. On both plots, $\delta=0.15$,
%$\beta=40$, system size $128 \times128$. In (a), $U=5.75$ (b) $U=6.25$.}} }%
%BeginExpansion
\caption
{(Color) Fermi surface plots obtained from energy dispersion curves integrated from
$-0.2$ to $0.1$ with $|t'|=0.175$, $|t''|=0.05$. On both plots, $\delta=0.15$,
$\beta=40$, system size $128 \times128$. In (a), $U=5.75$ (b) $U=6.25$.}
%EndExpansion
%TCIMACRO{\TeXButton{fig4}{\label{fig4}} }%
%BeginExpansion
\label{fig4}
%EndExpansion
%TCIMACRO{\TeXButton{End Figure}{\end{figure}}}%
%BeginExpansion
\end{figure}%
%EndExpansion
Fig.\ 4 shows Fermi surface maps for all wave vectors $\mathbf{k}$ in the
first quadrant of the Brillouin zone. The maps are obtained, as in ARPES
experiments on NCCO\ \cite{Armitage:2001}, from the integral of the
single-particle spectral weight $A\left(  \mathbf{k},\omega\right)  $ times
the Fermi function over a frequency range running from $-0.2$ to $+0.1$. For
$U=5.75$, shown on the left-hand side, two hot spots are clearly apparent at
the intersection of the Fermi surface with the AFM zone boundary, as observed
experimentally at optimal doping. The AFM correlation length $\xi$ is $12$
lattice spacings for this plot and the spin susceptibility at $\left(  \pi
,\pi\right)  $ is much larger than the non-interacting value. At this $\beta$
and for this value of $U,$ a pseudogap is observed only at the hot spots. They
appear because the strong low-energy AFM fluctuations can scatter excitations
at these points to other points on the Fermi surface separated by $\left(
\pi,\pi\right)  \;$\cite{Preosti:1999}. If $U$ is not large enough, there is
only a decrease of spectral weight at the hot spots instead of a real
pseudogap. By contrast, the right-hand side of Fig.\ 4 shows that if the
interaction is too large, $U=6.25$ $\left(  \xi=18\right)  $, the AFM
fluctuations scatter so strongly that a pseudogap appears everywhere along an
arc on the Fermi surface. This confirms our contention that $U$ cannot be too
large near optimal doping in e-HTSC to explain the experimental results. The
value of $U,$ however, does have to increase with decreasing $\delta$ so as to
recover the Fermi maps observed at $\delta=10\%$ as well as the Mott insulator
at half-filling\ \cite{Bansil:2002,Senechal:2003}.

In summary, in e-HTSC the symmetry of the superconducting order parameter, the
dependence of $T_{c}$ on pressure, as well as the hot spots observed by ARPES
at optimal doping can all be explained by the Hubbard model at weak to
intermediate coupling. Generally, antiferromagnetic fluctuations help
superconductivity until they are so strong that they open up a pseudogap that
hinders $d-SC$.

We are especially grateful to V. Hankevych for discussions and for performing
some of the calculations. We also thank S. Allen, C. Bourbonnais, A.-M.
Dar\'{e}, P. Fournier and D. S\'{e}n\'{e}chal for useful discussions. This
work was partially supported by the Natural Sciences and Engineering Research
Council of Canada (NSERC), by the Fonds pour la Formation de Chercheurs et
l'Aide \`{a} la Recherche (FCAR) from the Qu\'{e}bec government, by the
Canadian Institute for Advanced Research and by the Tier I Canada Research
Chair program (A.-M.S.T).

\end{document}